\documentclass[prl,twocolumn,floatix,amssymb,showpacs,amsmath,superscriptaddress]{revtex4-1}
\pdfoutput=1
\usepackage{graphicx}
\usepackage{epsfig}
\usepackage{amsmath,amssymb}
\usepackage{ dsfont }
\usepackage{color}

\DeclareMathOperator{\Tr}{Tr}
\begin{document}

\newcommand{\ket}[1]{\ensuremath{\left|{#1}\right\rangle}}
\newcommand{\bra}[1]{\ensuremath{\left\langle{#1}\right|}}
\newcommand{\quadr}[1]{\ensuremath{{\not}{#1}}}
\newcommand{\quadrd}[0]{\ensuremath{{\not}{\partial}}}
\newcommand{\slpar}{\partial\!\!\!/}
\newcommand{\gtrescero}{\gamma_{(3)}^0}
\newcommand{\gtresuno}{\gamma_{(3)}^1}
\newcommand{\gtresi}{\gamma_{(3)}^i}

\title{Digital Quantum Simulation of Spin Systems in Superconducting Circuits}

\date{\today}

\author{U. Las Heras}
\affiliation{Department of Physical Chemistry, University of the Basque Country UPV/EHU, Apartado 644, 48080 Bilbao, Spain}
\author{A. Mezzacapo}
\affiliation{Department of Physical Chemistry, University of the Basque Country UPV/EHU, Apartado 644, 48080 Bilbao, Spain}
\author{L. Lamata}
\affiliation{Department of Physical Chemistry, University of the Basque Country UPV/EHU, Apartado 644, 48080 Bilbao, Spain}
\author{S. Filipp}
\affiliation{Department of Physics, ETH Z\"urich, CH-8093 Z\"urich, Switzerland}
\author{A. Wallraff}
\affiliation{Department of Physics, ETH Z\"urich, CH-8093 Z\"urich, Switzerland}
\author{E. Solano} 
\affiliation{Department of Physical Chemistry, University of the Basque Country UPV/EHU, Apartado 644, 48080 Bilbao, Spain}
\affiliation{IKERBASQUE, Basque Foundation for Science, Alameda Urquijo 36, 48011 Bilbao, Spain}

\begin{abstract}
We propose the implementation of a digital quantum simulator for prototypical spin models in a circuit quantum electrodynamics architecture. We consider the feasibility of the quantum simulation of Heisenberg and frustrated Ising models in transmon qubits coupled to coplanar waveguide microwave resonators. Furthermore, we analyze the time evolution of these models and compare the ideal spin dynamics with a realistic version of the proposed quantum simulator. Finally, we discuss the key steps for developing a toolbox of digital quantum simulators in superconducting circuits. \end{abstract}


\maketitle
The quantum coherent control of superconducting qubits has improved dramatically in the last years~\cite{Devoret13}. In this sense, circuit quantum electrodynamics (cQED)~\cite{Wallraff04} is considered as a potential scalable platform for quantum computing. Basic quantum algorithms~\cite{Fedorov12} and tests of fundamentals in quantum mechanics~\cite{Abdumalikov13} have been already realized. Single and two qubit gates~\cite{Chow12}, preparation of complex entangled states~\cite{Neeley10}, and basic protocols for quantum error correction~\cite{Reed12}, are among the quantum information tasks that can be performed with good fidelities. Moreover, superconducting circuits have reached sufficient complexity and potential scalability to be considered as quantum simulators.

A quantum simulator is a platform that allows us to reproduce the behavior of another quantum system. The original idea of quantum simulation can be traced back to Feynman~{\cite{Feynman82}, while the first mathematical formulation using local interactions was proposed some years later~\cite{Lloyd96}. So far, initial steps for quantum simulations in circuit QED have been taken, where a few analog quantum simulators have been proposed in superconducting qubits~\cite{Ripoll08,Tian10,Pritchett,Zhang13,Mei13,Ballester12,Pedernales13,Viehmann13}. On the other hand, an experiment  of discrete-time gate sequences to reproduce the dynamics of a given spin Hamiltonian has been recently realized in ion-trap~\cite{Lanyon11} and photonic~\cite{LanyonQChem} systems, together with proposals for the emulation of interacting fermionic-bosonic models~\cite{Casanova12,Mezzacapo12}. The digital decomposition of Hamiltonians and their implementation using short-time gates has been demonstrated to be efficient~\cite{Suzuki90,Berry07}. Accordingly, it is timely to address the topic of digital quantum simulators with superconducting circuits. The quantum simulation of spin models can shed light onto a variety of open problems, such as quantum phase transitions~\cite{Chen07}, correlated one-dimensional systems~\cite{Santos12}, and high-$T_c$ superconductivity~\cite{Anderson13}.

In this Letter, we investigate the possible implementation of digital quantum simulations of spin Hamiltonians in a superconducting setup consisting of several superconducting qubits coupled to a coplanar waveguide resonator. Although our proposal is valid for every superconductor-based qubit with long enough coherence time, we focus on a transmon qubit setup. Superconducting transmon qubits are commonly used because of their low sensitivity to offset charge fluctuations~\cite{Koch07}. However, depending on the targeted physics, other superconducting qubits may be considered for quantum simulations or quantum information processing. First, we show that a variety of spin dynamics can be retrieved by a digital decomposition in a generic quantum simulator. Then, we consider prototypical spin models, simulation times, and fidelities with current circuit QED technology, showing the computational power of superconducting qubits in terms of digital quantum simulations. In this way, we analyze the required resources in a realistic setup for a multipurpose quantum simulator of spin dynamics capable of emulating a general many-qubit spin Hamiltonian.

Most physical Hamiltonians can be written as a sum of local terms, $H=\sum_{k=1}^N H_k$, where each $H_k$ acts on a local Hilbert space. The dynamics of a generic Hamiltonian $H$ can be approximated by discrete stepwise unitaries, up to arbitrary small errors, according to the formula ($\hbar=1$ here and in the following)~\cite{Lloyd96},
\begin{eqnarray}
e^{-iHt} = && \left(e^{-iH_1t/l} \cdots e^{-iH_Nt/l}\right)^l \nonumber \\
 && + \sum_{i<j}  \frac{[H_i,H_j]t^2}{2l} + \sum_{k=3}^{\infty} E(k), \label{TrotterBasic}
\end{eqnarray} 
with $l||Ht/l ||_{\textrm{sup}}^k/k! \geq ||E(k)||_{\textrm{sup}}$ being an upper bound on the higher order error terms. In the trivial case, when $[H_i,H_j]=0$ for every $\{i,j\}$, the error made in the digital approximation is zero.
To approximate $e^{-iHt}$ to arbitrary precision, one can divide the simulated time $t$ into $l$ intervals of size $t/l$, and apply sequentially the evolution operator of each local term for every interval. Repeating the sequence $l$ times, the error can be made as small as desired just by increasing $l$. However, in a realistic quantum simulator, there will be a limit to the number of local $e^{-iH_kt/l}$ gates feasible to apply, due to accumulated gate errors. Accordingly, one has to optimize the number of steps $l$ to get the best possible result.

{\it Heisenberg interaction.}--- 
Digital methods can be used to simulate the Heisenberg spin model with available resources in superconducting circuits. We consider a setup made of several transmon qubits coupled to a single coplanar microwave resonator~\cite{Koch07},
\begin{eqnarray}
H^T=&&\omega_ra^{\dagger}a+\sum_{i=1}^N\Big{[}4E_{C,i}(n_i-n_{g,i})^2-E_{J,i}\cos\phi_i\nonumber\\
&&+2\beta_i eV_{\textrm{rms}}n_i(a+a^{\dagger})\Big{]}.\label{HT}
\end{eqnarray}
Here, $n_i$, $n_{g,i}$ and $\phi_i$ stand, respectively, for the quantized charge on the superconducting island, the offset charge and the quantized flux of the $i$th transmon qubit. The operators $a$($a^{\dagger}$) act on the resonator field, whose first mode has frequency $\omega_r$. $E_{C,i}$ is the charging energy of the superconducting island, while $E_{J,i}=E_{J,i}^{\textrm{max}}|\cos(\pi\Phi_i/\Phi_0)|$ is the Josephson energy of the dc-SQUID loop embedded in the $i$th qubit. The latter can be tuned from small values up to $E_{J,i}^{\textrm{max}}$ by changing the ratio between the external magnetic flux $\Phi_i$, that threads the loop, and the elementary flux quantum $\Phi_0$. Here, $\beta_i$ are renormalization coefficients of the couplings due to circuit capacitances, $V_{\textrm{rms}}$ is the root mean square voltage of the resonator, and $e$ is the electron charge. Typical transmon regimes consider ratios of Josephson to charging energy $E_J/E_C\gtrsim20$. 

Notice that cavity and circuit QED platforms do not feature the Heisenberg interaction from first principles. Nevertheless, one can consider a digital simulation of the model.
We show that the coupled transmon-resonator system, governed by the Hamiltonian in Eq.~(\ref{HT}), can simulate Heisenberg interactions of $N$ qubits, which in the case of homogeneous couplings reads
\begin{equation}
H^H=\sum_{i=1}^{N-1}  J\left(\sigma^x_i \sigma^x_{i+1} +\sigma^y_i \sigma^y_{i+1} +\sigma^z_i \sigma^z_{i+1}\right).\label{XYZinhom}
\end{equation}
Here the Pauli matrices $\sigma^j_i$, $j\in\{x,y,z\}$ refer to the subspace spanned by the first two levels of the $i$th transmon qubit. We begin by considering the simplest case, in which two qubits are involved. The $XY$ exchange interaction can be directly reproduced by dispersively coupling two transmon qubits to the same resonator~\cite{Blais04,Majer07,Filipp11}, $H_{12}^{xy}=  J \left( {\sigma_1^+} {\sigma_2 ^-}+  {\sigma_1^-} {\sigma_2^+} \right) =   J/2  \left(\sigma^x_1 \sigma^x_2 +  \sigma^y_1 \sigma^y_2\right)$.
The $XY$ exchange interaction can be transformed via local rotations of the single qubits to get the effective Hamiltonians 
$H_{12}^{xz} = R^x_{12}(\pi/4)H_{12}^{xy}R^{x \dagger}_{12}(\pi /4) = J/2  \left(\sigma^x_1 \sigma^x_2 +  \sigma^z_1 \sigma^z_2 \right)$ and $H_{12}^{yz}=R^{y}_{12}(\pi/4)H_{12}^{xy}R^{y \dagger}_{12}(\pi /4)= J/2  \left(\sigma^y_1 \sigma^y_2 +  \sigma^z_1 \sigma^z_2 \right)$. Here, $R_{12}^{x(y)}(\pi/4)=\exp[-i\pi/4(\sigma_1^{x(y)}+\sigma_2^{x(y)})]$ represents a local rotation of the first and second transmon qubits along the $x(y)$ axis.
The $XYZ$ exchange Hamiltonian $H_{12}^{xyz}$ can therefore be implemented according to the protocol shown in Fig.~\ref{Fig1}(a) with the following steps. {\it Step 1.---} The qubits interact for a time $t$ according to the $XY$ Hamiltonian $H_{12}^{xy}$.
{\it Step 2.---} Application of single qubit rotations $R^x_{12}(\pi/4)$ to both qubits. 
{\it Step 3.---} The qubits interact for a time $t$ with the $H_{12}^{xy}$ Hamiltonian.
{\it Step 4.---} Application of single qubit rotation $R^{x \dagger}_{12}(\pi/4)$ to both qubits. 
{\it Step 5.---} Application of single qubit rotation $R^y_{12}(\pi/4)$ to both qubits. 
{\it Step 6.---} The qubits interact for a time $t$ according to the $H_{12}^{xy}$ Hamiltonian.
{\it Step 7.---} Application of single qubit rotation $R^{y \dagger}_{12}(\pi/4)$ to both qubits. 
Consequently, the total unitary evolution reads
\begin{eqnarray}
U_{12}^H (t)&=& e^{-i H_{12}^{xy} t} e^{-i H_{12}^{xz} t} e^{-i H_{12}^{yz} t}= e^{-i H^{\textrm{H}}_{12}t}.\label{eq8}
\end{eqnarray}
This evolution operator simulates the dynamics of Eq.~(\ref{XYZinhom}) for two qubits. Arbitrary inhomogeneities of the couplings can be achieved by implementing different simulated phases for different digital steps. Notice that, in this case, just one Trotter step is needed to achieve a simulation without digital errors, due to the commutativity of $H_{12}^{xy}$, $H_{12}^{xz}$, and $H_{12}^{yz}$. Thus, from a practical point of view, the only source of errors will come from accumulated gate errors. One can assume two-qubit gates with an error of about $5\%$ and eight $\pi /4$ single qubit rotations with errors of $1\%$. This will give a total fidelity of the protocol around $77\%$. Moreover, the total execution time for a $\pi/4$ simulated $XYZ$ phase will be of about $0.10$~$\mu$s. Throughout the Letter, we compute the execution times by summing the corresponding times of all the employed gates, where we consider typical circuit QED values.
\begin{figure}
\includegraphics[scale=0.33]{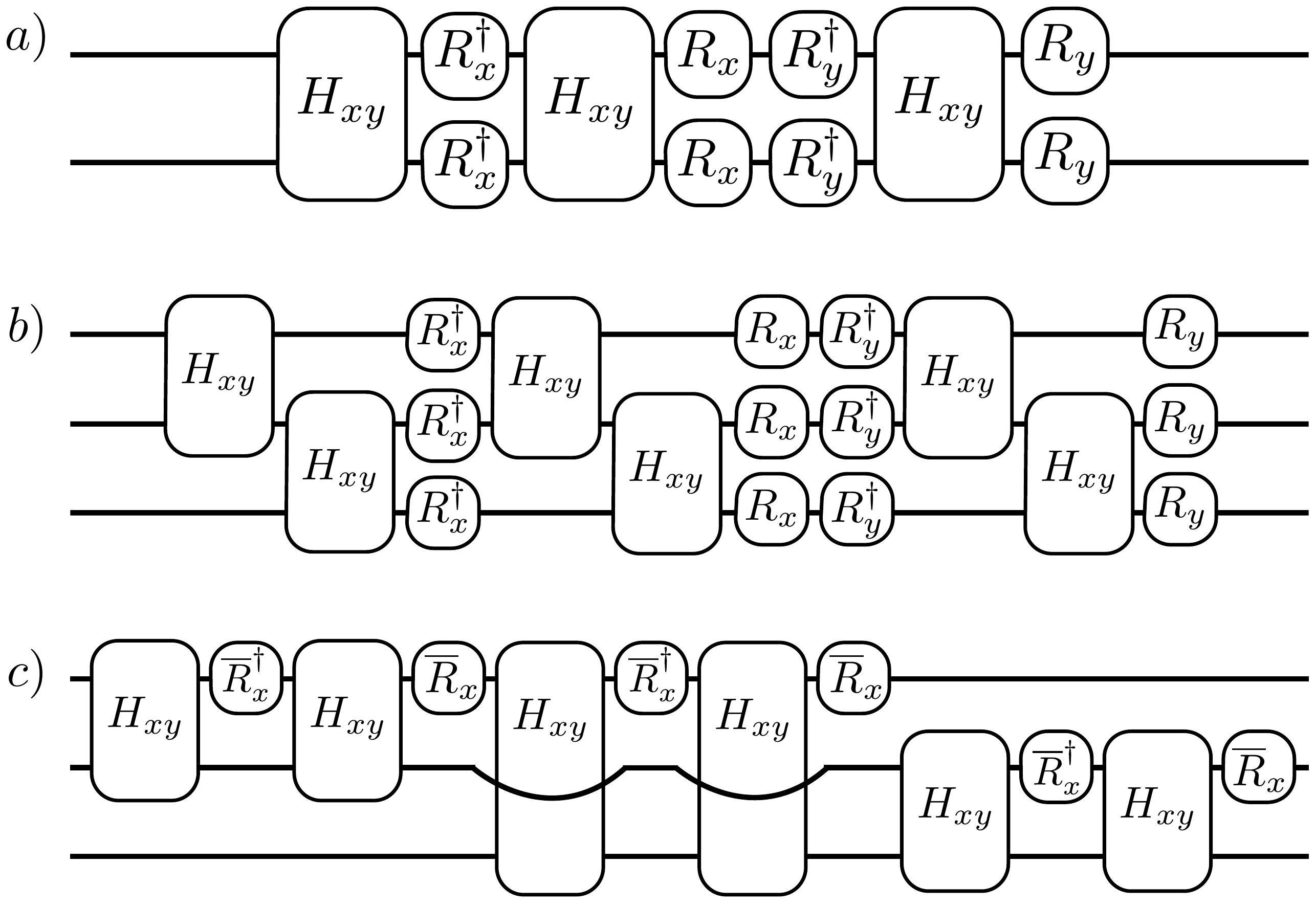}
\caption{Protocols for digital quantum simulations with transmon qubits. (a) Heisenberg model of two qubits. (b) Heisenberg model of three qubits. (c) Frustrated Ising model of three qubits. Here, $R_{x(y)}\equiv R^{x(y)}(\pi/4)$ and $\overline{R}_x\equiv R^x(\pi/2)$. Note that exchanging each $R$ matrix with its adjoint does not affect the protocols.\label{Fig1}} 
\end{figure} 

Now, we consider a digital protocol for the simulation of the Heisenberg interaction for a chain of three spins. When considering more than two spins, one has to take into account noncommuting Hamiltonian steps, involving digital errors. This three-spin case is directly extendable to arbitrary numbers of spins. We follow a digital approach for its implementation, as shown in Fig.~\ref{Fig1}(b). {\it Step~1.---}~Qubits 1 and 2 interact for a time $t/l$ with the $XY$ Hamiltonian. {\it Step~2.---}~Qubits 2 and 3 interact for a time $t/l$ with the $XY$ Hamiltonian. {\it Step~3.--}~Application of $R^x_{12}(\pi/4)$ to each qubit.
{\it Step~4.---}~Qubits 1 and 2 interact for a time $t/l$ with the $XY$ Hamiltonian.
{\it Step~5.---}~Qubits 2 and 3 interact for a time $t/l$ with the $XY$ Hamiltonian.
{\it Step~6.---}~Application of $R^{x\dagger}_{12}(\pi/4)$ to each qubit. 
{\it Step~7.---}~Application of $R^y_{12}(\pi/4)$ to each qubit. 
{\it Step~8.---}~Qubits 1 and 2 interact for a time $t/l$ with the $XY$ Hamiltonian.
{\it Step~9.---}~Qubits 2 and 3 interact for a time $t/l$ with the $XY$ Hamiltonian.
{\it Step~10.---}~Application of $R^{y\dagger}_{12}(\pi/4)$ to each qubit.
Thus, the total unitary evolution per step reads
\begin{eqnarray}
U_{123}^{H}(t/l) =&&e^{-i H_{12}^{xy} t/l}e^{-i H_{23}^{xy} t/l} e^{-i H_{12}^{xz} t/l}\nonumber e^{-i H_{23}^{xz} t/l}\\
&&\times\ e^{-i H_{12}^{yz} t/l}e^{-i H_{23}^{yz} t/l} \label{eq15}.
\end{eqnarray}
\begin{figure}
\includegraphics[scale=0.33]{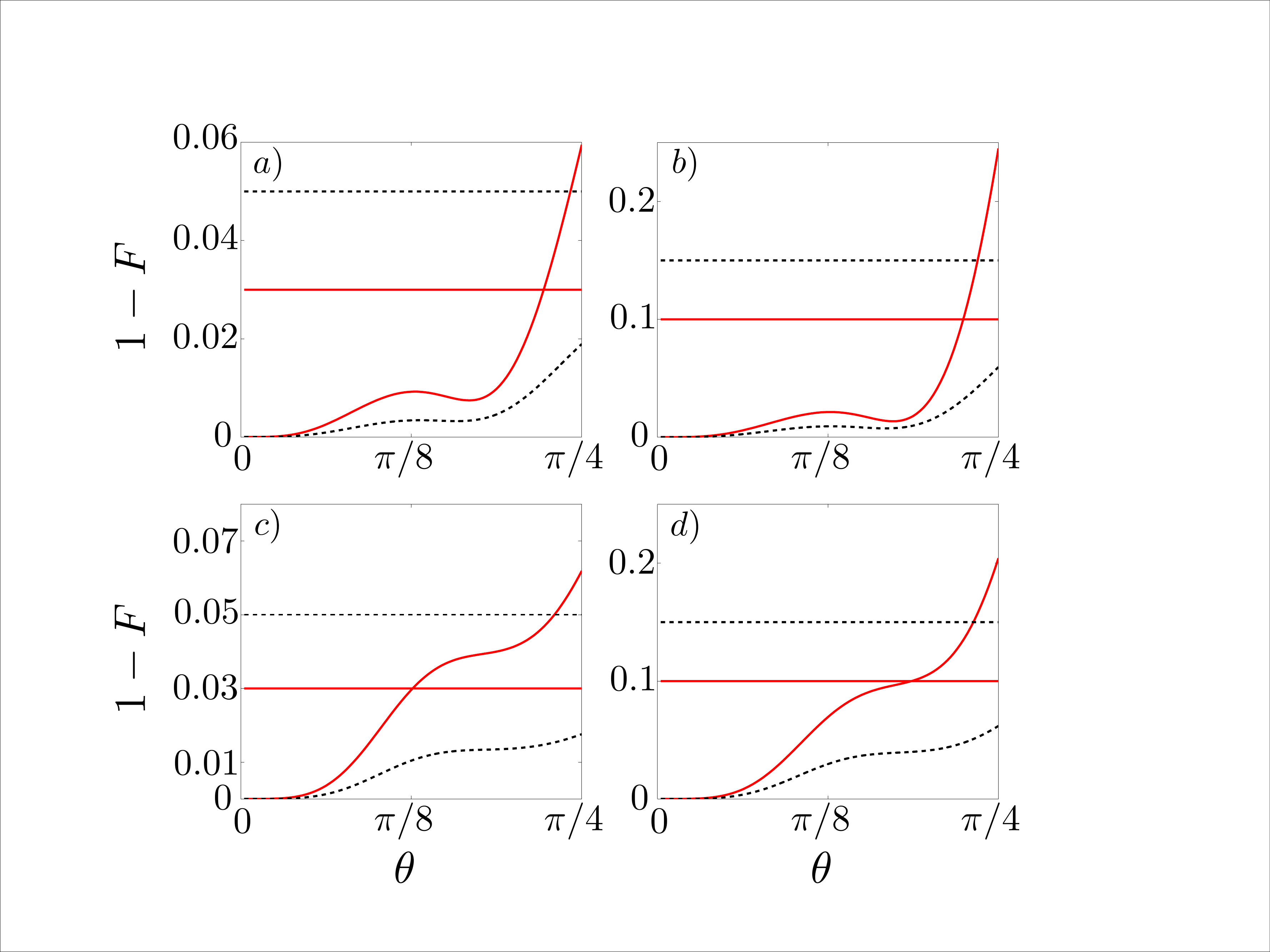}
\caption{ Fidelity loss for simulated Hamiltonians for three qubits, in the interval $\theta=[0,\pi/4]$, $\theta\equiv Jt$. Curved lines show digital errors, while horizontal lines show the accumulated error due to a single step error of $\epsilon$. Red solid (black dotted) lines stand for lower (higher) digital approximations $l$. (a) Heisenberg model, with $\epsilon=10^{-2}$, $l=3,5$, and (b) $\epsilon=5\times10^{-2}$, $l=2,3$. (c) Transverse field Ising model, with $\epsilon=10^{-2}$, $l=3,5$ and (d) $\epsilon=5\times10^{-2}$, $l=2,3$.\label{Fig2}}\label{Fig2}
\end{figure}
In this case, the protocol has to be repeated $l$ times according to Eq.~(\ref{TrotterBasic}), to approximate the dynamics of Eq.~(\ref{XYZinhom}) for three qubits. Each Trotter step involves four single qubit gates at different times and six two qubit gates, producing a step time of about $0.16$~$\mu$s, which is well below standard coherence times for transmon qubits~\cite{Rigetti12}. In Figs.~\ref{Fig2}(a) and \ref{Fig2}(b), we plot the digital error of the simulated Heisenberg model for three qubits, along with horizontal lines, that show the error of the imperfect gates multiplied by the number of Trotter steps, i.e., the total accumulated gate error. In this way, one can distinguish time domains dominated by the digital error and time domains in which the largest part of the error in the quantum simulation is due to experimental gate errors. One can consider interactions with open and closed boundary conditions, adding an extra term coupling the first and last spin. Extending this protocol to $N$ qubits with open or periodic boundary conditions, we compute an upper bound on the second order Trotter error $E_\textrm{open}=24(N-2)(Jt)^2/l$ and $E_\textrm{periodic}=24N(Jt)^2/l$.

{\it Ising interaction.--- }
Here, we consider a generic $N$ qubit Ising interaction $J\sum_i\sigma_i^x\sigma_{i+1}^x$, with periodic boundary conditions. Considering a three site model is sufficient to show the effect of frustration in the system. The antiferromagnetic interaction is inefficiently solvable in a classical computer, while it is efficient for a quantum simulator~\cite{KimFrustration}. We consider the isotropic antiferromagnetic case between three sites , $H^I_{123} =  J \sum_{i<j}\sigma_i^x \sigma_j^x$, with $i,j=1,2,3$ and $J>0$. In order to simulate this Hamiltonian, one can apply a $\pi/2$ rotation to one of the qubits. This will result in an effective stepwise elimination of the $YY$ component of interaction,
\begin{equation}
H_{12}^{x-y}={R}^{x}_1(\pi/2)H_{12}^{xy}{R}_1^{x\dagger}(\pi /2) = J \left(\sigma^x_1 \sigma^x_2 -  \sigma^y_1 \sigma^y_2 \right).
\end{equation}
The protocol for the simulation is shown in Fig.~\ref{Fig1}(c).
As the terms of the Ising Hamiltonian commute, there is no error from the Trotter expansion. We obtain a fidelity of the protocol of about $64\%$. The time for the execution of all gates is $0.18$~$\mu$s.                                                                                                                            

\begin{table}
\caption{Execution times and error bounds for the Heisenberg($H$) and Ising($I$) models with open($o$) and periodic($p$) boundary conditions for $N$ qubits. Here $\theta\equiv Jt$, $J/2$ and $g_\phi$ are, respectively, the coupling strength of the $XY$ and single-qubit gates, and $\tau_s$ is the pulse time required for a single qubit rotation. \label{Table}}
\vspace{0.2cm}
\begin{tabular}{l|l|l}
\hline\hline

  			& Execution time & Error bound\\
\hline
  $H_{o}$ & $4l\tau_s+6(N-1)\theta/J$& $24(N-2)(Jt)^2/l$\\
  $ H_{p}$  & $4l\tau_s+6N\theta/J$  & $24N(Jt)^2/l$\\
  $I_{o}$ & $2(N-1)l\tau_s+\theta/g_\phi+4(N-1)\theta/J$ & $2(N-1)(Jt)^2/l$\\
  $I_{p}$  & $2Nl\tau_s+\theta/g_\phi+4N\theta/J$ & $2N(Jt)^2/l$\\

\hline\hline
\end{tabular}
\end{table}
One can also add a transverse magnetic field, that leads to the Hamiltonian $H^{IT}_{123} =  J \sum_{i<j}\sigma^x_i \sigma^x_j + B \sum_i \sigma^y_i$. In this case, the terms of the Hamiltonian do not commute, so we need to apply more than one Trotter step to achieve adequate fidelities. The unitary evolution per Trotter step in this case is given by
\begin{eqnarray}
U(t/l)=&&\ e^{-i H_{12}^{xy} t/l}e^{-i H_{12}^{x-y} t/l}e^{-i H_{13}^{xy}t/l} e^{-i H_{13}^{x-y}t/l}\nonumber\\
&&\ \times\ e^{-i H_{23}^{xy}t/l}e^{-i H_{23}^{x-y}t/l}e^{-i  Bt/l(\sigma^y_1+\sigma^y_2+\sigma^y_3)}\nonumber\\
=&&\ e^{-i  2Jt/l(\sigma^x_1 \sigma^x_2 + \sigma^x_1 \sigma^x_3 + \sigma^x_2 \sigma^x_3) }e^{-i  Bt/l(\sigma^y_1+\sigma^y_2+\sigma^y_3)}.\nonumber\\ \label{eq20}
\end{eqnarray}
In Figs.~\ref{Fig2}(c) and \ref{Fig2}(d), we plot the fidelity loss for different numbers of Trotter steps, in the three-qubit frustrated Ising model with transverse magnetic field, considering an error for each step due to the imperfect gates. The time for simulating the transverse field Ising model for the considered dynamics is about 190 ns per Trotter step. The protocol can also be extended to $N$ qubits with open and periodic boundary conditions, where we compute an upper bound to the second order error in $Jt/l$ of $E_\textrm{open}=2(N-1)(Jt)^2/l$ and $E_\textrm{periodic}=2N(Jt)^2/l$. We report in Table~\ref{Table} execution times and error bounds for the models proposed, for $N$ qubits. In general, given the nonlocal character of the microwave resonator acting as a quantum bus, one can emulate 2D and 3D interaction topologies.

In order to estimate the feasibility of the protocols in a superconducting circuit setup, we perform a numerical simulation for the Heisenberg interaction between two transmon qubits coupled to a coplanar waveguide resonator. We compute the effect on the protocol of a realistic $XY$ interaction, given as an effective second order Hamiltonian, obtained from the first order Hamiltonian,
\begin{align}
H_{\textrm t}=&\sum_{i=0}^2\sum_{j=1}^2\omega_i^j \ket{i,j}\bra{i,j}+\omega_ra^{\dagger}a\nonumber\\
&+\sum_{i=0}^2\sum_{j=1}^2 g_{i,i+1}(\ket{i,j}\bra{i+1,j}+\textrm{H.c.})(a+a^{\dagger}).\label{HamcircuitQED}
\end{align}

Here, $\omega_i^j$ is the transition energy of the $i$th level, with respect to the ground state, of the $j$th qubit, and $\omega_r$ is the transition frequency of the resonator. We consider the first three levels for each transmon qubit, and a relative anharmonicity factor of $\alpha_r=(\omega_2^j-2\omega_1^j)/\omega_1^j=-0.1$, typical for the transmon regime~\cite{Koch07}. We assume identical transmon devices, with transition frequencies $\omega_1^{1,2}\equiv \omega_1=2\pi\times5$~GHz. The resonator frequency is set to $\omega_r=2\pi\times7.5$~GHz. We consider the coupling between different levels of a single transmon qubit~\cite{Koch07} $g_{i,i+1}=\sqrt{i+1}g_0$, where $g_0=2\beta eV_{\textrm{rms}}=2\pi\times200$~MHz. The chosen experimental parameters are typical for superconducting circuit setups and they can be optimized for each platform. The resonator-transmon coupling Hamiltonian, in the interaction picture with the free energy $\sum_{i,j}\omega_i^j \ket{i,j}\bra{i,j}+\omega_ra^{\dagger}a$, results in an effective coupling between the first two levels of the two transmon qubits $H_{\textrm{eff}}=[g_{01}^2\omega_1/(\omega_1^2-\omega_r^2)](\sigma_1^x\sigma_2^x+\sigma_1^y\sigma_2^y)$, where we have considered negligible cavity population $\langle a^{\dagger}a\rangle\approx0$ and renormalization of the qubit frequencies to cancel Lamb shifts. Here we have defined a set of Pauli matrices for the subspace spanned by the first two levels of each transmon, e.g. $\sigma_{1(2)}^x\equiv\ket{0,1(2)}\bra{1,1(2)}+\textrm{H.c.}$
In order to estimate the effect of decoherence in a realistic setup, we consider the master equation dynamics,
\begin{equation}
\dot{\rho}=-i[H_{\textrm{t}},\rho]+\kappa L(a)\rho+\sum_{i=1}^2\left(\Gamma_\phi L(\sigma^z_i)\rho+\Gamma_- L(\sigma_i^-)\rho\right),\label{master}
\end{equation}
where we have defined the Lindblad superoperators $L(\hat{A})\rho=(2\hat{A}\rho \hat{A}^{\dagger}-\hat{A}^{\dagger}\hat{A}\rho-\rho \hat{A}^{\dagger}\hat{A})/2$. We have set a decay rate for the resonator of $\kappa=2\pi\times10$~kHz, and a dephasing and decay rate for the single transmon qubit of $\Gamma_\phi=\Gamma_-=2\pi\times20$~kHz. We perform a numerical simulation for the Heisenberg protocol for two transmon qubits, following the steps as in Fig.~\ref{Fig1}(a), using for the $XY$ interaction steps the result of the dynamics obtained by solving Eq.~(\ref{master}), and ideal single-qubit rotations. The result is plotted in Fig.~\ref{NumSim}. The evolution for the density matrix $\rho$, that encodes the dynamics of the two transmon qubits, is compared to the exact quantum evolution $|\Psi\rangle_I$, that evolves according to the Hamiltonian in Eq.~(\ref{XYZinhom}), with $J=g_{01}^2\omega_1/(\omega_1^2-\omega_r^2)\approx2\pi\times6$~MHz. One can observe that good simulation fidelities $F=\Tr(\rho|\Psi_I\rangle\langle\Psi_I|)$ are achieved for nontrivial dynamics. Note that the action of the Heisenberg Hamiltonian on an initial state, which is also an eigenstate of the $\sigma_1^z\sigma_2^z$ operator, would be equivalent to the one of the $XY$ exchange interaction. To show signatures of the Heisenberg interaction, we choose in our simulation an initial state which does not have this property. One can also notice the typical small time-scale fidelity oscillations  due to the first order part of the dispersive exchange interaction. By further detuning the qubits from the resonator, one can reduce the contribution of the nondispersive part of the interaction, and increase the global fidelity of the protocol.  
\begin{figure}
\includegraphics[scale=0.3118]{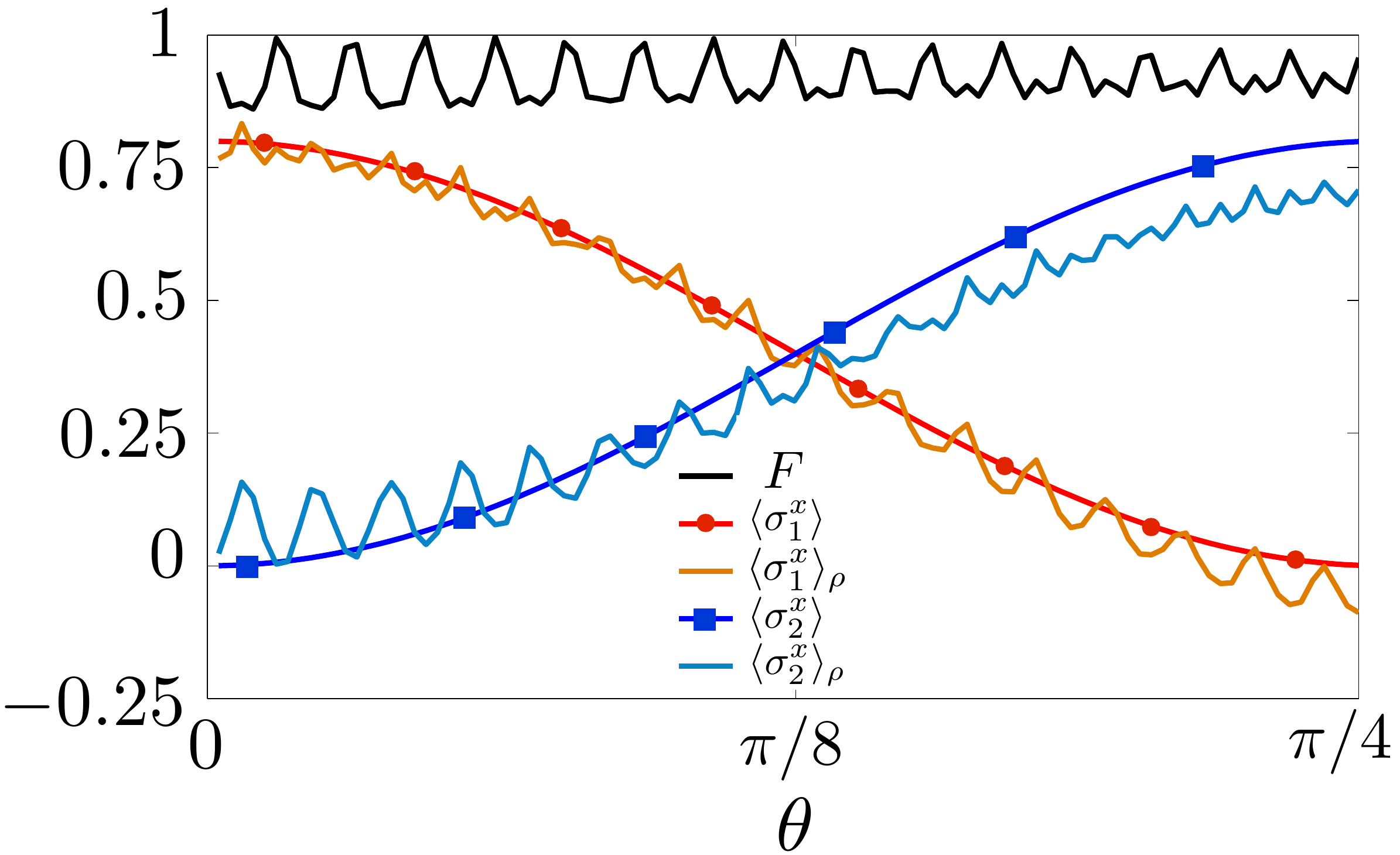}
\caption{ Dynamics for the simulated Heisenberg model for two transmon qubits, which are initialized in the state $1/\sqrt{5}(\ket{\uparrow}+2\ket{\downarrow})\otimes\ket{\downarrow}$. Fidelity $F=\Tr(\rho|\Psi_I\rangle\langle\Psi_I|)$ shows the behavior of the protocol for a given simulated phase $\theta$. The ideal spin dynamics $\langle\sigma_i^x\rangle$ for both qubits is plotted versus mean values $\langle\sigma_{i}^x\rangle_\rho$ obtained with the qubit Hamiltonian $H_{t}$.\label{NumSim}} 
\end{figure}

In conclusion, we have proposed a digital quantum simulation of spin chain models in superconducting circuits. We have considered prototypical models such as the Heisenberg and frustrated Ising interactions. Furthermore, we have shown the feasibility of the simulation with state-of-the-art technology of transmon qubits coupled to microwave resonators. In the near future, these protocols may be extended to many-qubit spin models, paving the way towards universal quantum simulation of spin dynamics in circuit QED setups. 

We acknowledge funding from the Basque Government IT472-10, Spanish MINECO FIS2012-36673-C03-02, Ram\'on y Cajal RYC-2012-11391, UPV/EHU UFI 11/55, CCQED,
PROMISCE and SCALEQIT EU projects.


\begin{thebibliography}{99}

\bibitem{Devoret13} M. H. Devoret and R. J. Schoelkopf, Science {\bf 339}, 1169 (2013).

\bibitem{Wallraff04}A. Wallraff, D. I. Schuster, A. Blais, L. Frunzio, R.-S. Huang, J. Majer, S. Kumar, S. M. Girvin, and R. J. Schoelkopf, Nature (London) {\bf 431}, 162 (2004).

\bibitem{Fedorov12}A. Fedorov, L. Steffen, M. Baur, M. P. da Silva, and A. Wallraff, Nature (London) {\bf 481}, 170 (2012).

\bibitem{Abdumalikov13} A. A. Abdumalikov, J. M. Fink, K. Juliusson, M. Perchal, S. Berger, A. Wallraff, and S. Filipp, Nature (London) {\bf 496}, 482 (2013).

\bibitem{Chow12} J. M. Chow, J. M. Gambetta, A. D. C\'orcoles, S. T. Merkel, J. A. Smolin, C. Rigetti, S. Poletto, G. A. Keefe, M. B. Rothwell, J. R. Rozen, M. B. Ketchen, and M. Steffen, Phys. Rev. Lett. {\bf 109}, 060501 (2012).

\bibitem{Neeley10} M. Neeley, R. C. Bialczak, M. Lenander, E. Lucero, M. Mariantoni, A. D. O'Connell, D. Sank, H. Wang, M. Weides, J. Wenner, Y. Yin, T. Yamamoto, A. N. Cleland, and J. M. Martinis, Nature (London) {\bf467}, 570 (2010).

\bibitem{Reed12} M. D. Reed, L. DiCarlo, S. E. Nigg, L. Sun, L. Frunzio, S. M. Girvin, and R. J. Schoelkopf, Nature (London) {\bf 482}, 382 (2012).

\bibitem{Feynman82}R. P. Feynman, Int. J. Theor. Phys. {\bf 21}, 467 (1982).

\bibitem{Lloyd96}S. Lloyd,  Science {\bf 273}, 1073 (1996).

\bibitem{Ripoll08}J. J. Garc\'{\i}a-Ripoll, E. Solano, and M. A. Martin-Delgado, Phys. Rev. B {\bf 77}, 024522 (2008).

\bibitem{Tian10}L. Tian, Phys. Rev. Lett. {\bf105}, 167001 (2010).

\bibitem{Pritchett} 	
E. J. Pritchett, C.  Benjamin, A. Galiautdinov, M. R. Geller, A. T. Sornborger, P. C. Stancil, and J. M. Martinis, arXiv:1008.0701.

\bibitem{Zhang13}Y. Zhang, L. Yu, J.-Q. Liang, G. Chen, S. Jia, and F. Nori, Sci. Rep. {\bf 4}, 4083 (2014).

\bibitem{Mei13} F. Mei, V. M. Stojanovi\'c, I. Siddiqi, and L. Tian, Phys. Rev. B {\bf 88}, 224502 (2013).

\bibitem{Ballester12} D. Ballester, G. Romero, J. J. Garc\'ia-Ripoll, F. Deppe, and E. Solano, Phys. Rev. X {\bf 2}, 021007 (2012).

\bibitem{Pedernales13} J. S. Pedernales, R. Di Candia, D. Ballester, and E. Solano, New. J. Phys. {\bf15}, 055008 (2013). 

\bibitem{Viehmann13}O. Viehmann, J. von Delft, and F. Marquardt, Phys. Rev. Lett. {\bf 110}, 030601 (2013).

\bibitem{Lanyon11} B. P. Lanyon, C. Hempel, D. Nigg, M. M\"uller, R.~Gerritsma, F. Z\"ahringer, P. Schindler, J. T. Barreiro, M.~Rambach, G. Kirchmair, M. Hennrich, P. Zoller, R.~Blatt, and C. F. Roos, Science {\bf 334}, 57 (2011).

 \bibitem{LanyonQChem} B. P. Lanyon, J. D. Whitfield, G. G. Gillet, M. E. Goggin, M. P. Almeida, I. Kassal, J. D. Biamonte, M. Mohseni, B. J. Powell, M. Barbieri, A. Aspuru-Guzik, and A. G. White, Nat. Chem. {\bf 2}, 106 (2009).

\bibitem{Casanova12}J. Casanova, A. Mezzacapo, L. Lamata, and E. Solano, Phys. Rev. Lett. {\bf 108}, 190502 (2012).

\bibitem{Mezzacapo12}A. Mezzacapo, J. Casanova, L. Lamata, and E. Solano, Phys. Rev. Lett. {\bf 109}, 200501 (2012).

\bibitem{Suzuki90} M. Suzuki, Phys. Lett. A {\bf 146}, 319 (1990).

\bibitem{Berry07} D. W. Berry, G. Ahokas, R. Cleve, and B. C. Sanders, Commun. Math. Phys. {\bf 270}, 359 (2007).

\bibitem{Chen07} S. Chen, L. Wang, S.-J. Gu, and Y. Wang, Phys. Rev. E {\bf 76}, 061108 (2007).

\bibitem{Santos12}L. F. Santos, F. Borgonovi, and F. M. Izrailev, Phys. Rev. Lett. {\bf 108}, 094102 (2012).

\bibitem{Anderson13}P. W. Anderson, Science {\bf 235}, 1196 (1987).

\bibitem{Koch07} J. Koch, T. M. Yu, J. Gambetta, A. A. Houck, D. I. Schuster, J. Majer, A Blais, M. H. Devoret, S. M. Girvin and R. J. Schoelkopf, Phys. Rev. A {\bf 76}, 042319 (2007).

\bibitem{Blais04}A. Blais, R.-S. Huang, A. Wallraff, S. M. Girvin, and R. J. Schoelkopf, Phys. Rev. A {\bf 69}, 062320 (2004).

\bibitem{Majer07}J. Majer, J. M. Chow, J. M. Gambetta, Jens Koch, B. R. Johnson, J. A. Schreier, L. Frunzio, D. I. Schuster, A. A. Houck, A. Wallraff, A. Blais, M. H. Devoret, S. M. Girvin, and R. J. Schoelkopf, Nature (London) {\bf 449}, 443 (2007).

\bibitem{Filipp11}S. Filipp, M. G\"oppl, J. M. Fink, M. Baur, R. Bianchetti, L. Steffen, and A. Wallraff, Phys. Rev. A {\bf 83}, 063827 (2011). 

\bibitem{Rigetti12}C. Rigetti, J. M. Gambetta, S. Poletto, B. L. T. Plourde, J. M. Chow, A. D. C\'orcoles, J. A. Smolin, S. T. Merkel, J. R. Rozen, G. A. Keefe, M. B. Rothwell, M. B. Ketchen, and M. Steffen, Phys. Rev. B {\bf 86}, 100506(R) (2012).
 
 \bibitem{KimFrustration} K. Kim, M.-S. Chang, S. Korenblit, R. Islam, E. E. Edwards, J. K. Freericks, G.-D. Lin, L.-M. Duan, and C. Monroe, Nature (London) {\bf 465}, 590 (2010).
\end{thebibliography}
\end{document}